\documentclass[12pt,preprint]{aastex}

\newcommand{\mgi}{Mg\,I}
\newcommand{\mgii}{Mg\,II}
\newcommand{\invcm}{cm$^{-1}$}
\newcommand{\mic}{\mu \mathrm m}

\shorttitle{On the Mg I lines at 12\,$\mu$m in Procyon}
\shortauthors{Ryde, Korn, Richter, \&\ Ryde}

\begin{document}


\title{The Zeeman-sensitive emission lines of Mg I at 12\,$\mu$m in Procyon}


\author{N. Ryde\altaffilmark{1,2}}
\affil{Department of Astronomy and Space Physics, Uppsala University, Box 515, SE-751\,20 Uppsala, Sweden}
\email{ryde@astro.uu.se}

\author{A. J. Korn}
\affil{Department of Astronomy and Space Physics, Uppsala University, Box 515, SE-751\,20 Uppsala, Sweden}
\email{akorn@astro.uu.se}

\author{M. J. Richter\altaffilmark{2}}
\affil{Department of Physics, University of California at Davis, CA 95616}
\email{richter@physics.ucdavis.edu}

\author{F. Ryde\altaffilmark{3}}
\affil{Stockholm Observatory, AlbaNova University Center, SE-106\,91 Stockholm, Sweden}
\email{felix@astro.su.se}


\altaffiltext{1}{Postdoctoral Fellow in 2001-2002 at the Department of Astronomy,
University of Texas at Austin, TX 78712}
\altaffiltext{2}{Visiting Astronomer at the Infrared Telescope Facility,
which is operated by the University of Hawaii under Cooperative Agreement
no. NCC 5-538 with the National Aeronautics and Space Administration, Office
of Space Science, Planetary Astronomy Program.}
\altaffiltext{3}{Postdoctoral Fellow in 2001-2002 at the Department of Physics,
Stanford University, CA 94305}


\begin{abstract}
Emission-lines of magnesium at $12\,\micron$ have been observed in the
spectrum of Procyon. We reproduce the observed, disk-averaged line flux from
Procyon (as well as the observed intensity profiles from the Sun) by calculating the line formation,
relaxing the assumption of Local Thermodynamic Equilibrium. We
find that the lines in Procyon are formed in the photosphere in
the same manner as the solar lines. We discuss our modeling of
these Rydberg lines and evaluate, among other things, the
importance of the ionizing flux and updated model-atom parameters. The
lines are of large diagnostic value for measurements of stellar
magnetic-fields through their Zeeman splitting.
We have not, however, detected splitting of the \mgi\
lines in Procyon.
Using simple arguments, we believe we would have detected
a magnetic field, had it been of a strength larger than approximately 800 Gauss covering more
than a quarter of the surface.
We discuss the prospects for future use of the Zeeman-sensitive, mid-infrared,
\mgi\ emission lines as a diagnostic tool for stellar magnetic fields.

\end{abstract}



\keywords{stars: individual ($\alpha$ CMi) --
             stars:  atmospheres--
             Infrared: stars}


\section{Introduction}

Although magnetic fields affect, and sometimes dominate,
the appearance of cool stars, direct detection and determination
of magnetic field strengths are difficult.
Methods involving polarized light and magnetograms are problematic
for cool stars with assumed solar-like magnetic geometries \citep{saar}.
On the other hand, studies of unpolarized Zeeman-splitting in the near-infrared
have given a large
scientific return \citep[see, for example, ][]{muglach92,solanki92,solanki:154,valenti95,johnskrull99}.

\mgi\ emission lines at $12~\,\micron$ in sunspot penumbrae
and plages clearly show Zeeman splitting and so trace the local
magnetic field \citep[see, for example,
][]{brault83,bruls:1,bruls:2}.
Zeeman splitting is independent of the field direction and will
show up even in a disk-averaged spectrum from a star with a
disorganized field structure, which is not the case for polarized
light.
The wavelength dependence of Zeeman broadening compared to
other types of broadening results in
the $12\,\micron$ lines
being more Zeeman-sensitive than near-infrared (NIR) lines. Furthermore, the lines
probe other atmospheric layers than the NIR lines.
Thus, these lines
potentially offer an excellent method
for measuring stellar disk-averaged magnetic fields.


Solar emission features at $12\,\micron$ were first mentioned in 1981
(Murcray et al. 1981), and attracted much attention during the 1980s.
In 1983, Chang \& Noyes \nocite{chang83} identified the strongest two
emission lines as coming from
transitions between Rydberg states in \mgi.
\citet{lemke87} showed that the emission could be formed in
the photosphere through stimulated emission and different departures
from Local Thermodynamic Equilibrium (LTE) for the lower and upper
levels. The first non-LTE modelling giving the lines in emission were by Carlsson et
al. (1990, 1992b) and Chang et al. (1991). The line
formation process was first explained successfully and in detail by
\citet{mc_mg}. They concluded that the lines were photospheric in origin and accounted
very well for their emission strengths and their complicated intensity
profiles as a function of the viewing angle on the solar disk.   \citet{uiten},
observing just one of the two lines, detected
emission from the K giant $\alpha$ Boo and saw
absorption from the five M giants and supergiants in their sample.  Their
model calculations were unable to fit the data.
\citet{ryde&richter} detected and analysed both Mg I emission lines
from Procyon ($\alpha$~CMi) observed at a resolving power of $R = 86\,000$.
As high-resolution, mid-infrared spectrographs become available on $8-10$~m
telescopes, these lines will be observed in many more stars.  A first
step towards the sophisticated use of these lines for measuring stellar
magnetic fields is the understanding of their formation and response to
a broad range of stellar parameters.

In this paper, we discuss our analysis of the
magnesium emission lines detected from Procyon.
We
discuss here the atomic model of \mgi\
and the assumptions made in our non-LTE calculation,
such as the collisional rates, the atmospheric radiation field
used, and the validity of the assumption of atmospheric homogeneity.
We conclude with a discussion on using these lines for determining
disk-averaged, magnetic-field strengths in stars and an estimate
of the maximum field strength and corresponding filling factor allowed
by our observations.


\section{Observations}

We observed Procyon with the Texas Echelon-Cross-Echelle
Spectrograph~\citep[TEXES,][]{texes}, a visitor instrument
at the NASA 3.0 meter Infrared Telescope Facility (IRTF) on Mauna Kea in Hawaii.
The observations took place over a number of observing runs from November
2000 through January 2004.
We used the high-resolution, cross-dispersed mode,
with orders covering $\sim 0.67$\,\invcm\
and a total spectral coverage of roughly 0.6\% (or $\sim 5\,$\,\invcm ).
Separate settings were required for each \mgi\ line.
Observations of narrow line sources indicate that the instrumental
line profile has a Gaussian core with a full-width at half maximum
(FWHM)$ = 3.5$~km\,s$^{-1}$, i.e.
$R\sim86\,000$, and wings that are only slightly stronger than a Gaussian
distribution.

Our observations consisted of calibration frames mixed with nodded
observations of Procyon.  Before each data file, we observed an ambient
temperature blackbody, low-emissivity surface, and the sky, as described
in~\citet{texes}.  The blackbody serves as a flat-field and provides
first-order sky correction after subtracting the sky emission frame.
We integrated for approximately 1 second per frame to overcome detector
read-out noise.  Depending on conditions, we summed 6 to 8 frames before
moving the telescope by 4 to 5 seconds of arc, thus placing Procyon at a
different position along the slit and creating a nod pair.
After roughly 10 minutes of integrating
on target, we started another series with the calibration frame.

We used the standard TEXES data reduction pipeline~\citep{texes} to reduce
the data.  We fixed spikes and cosmic ray hits using the time series,
rectified the echellogram so that the spatial and dispersion axes run
along rows and columns of the detector,
and removed the sky and telescope background by differencing each nod
pair.   The individual nod pairs were combined after allowing for as
much as a 1\arcsec\ shift along the slit and after weighting the pairs
according to the square of an estimated signal-to-noise ratio.  Extraction
of the final spectrum for a given data set used the spatial profile derived
from the data, with a mean set to zero, as the extraction template.
To set the frequency scale, we used the location and frequency of telluric
atmospheric lines, which typically give an accuracy better than 1~km\,s$^{-1}$.

After reducing a given data set, we normalized the continuum
using a 6$^{\rm th}$ order Legendre polynomial and then
combined appropriate data files.
Procyon has almost no photospheric features at this wavelength so
determining the continuum is straightforward and reliable.  The telluric atmosphere
at the \mgi\ wavelengths is very clean, so that we did not need to
correct for telluric features.
Slight variations in wavelength setting from night to night result in
increased noise where fewer data could be co-added.

In the observations presented in Figures \ref{em_811} and \ref{em_818},
the signal-to-noise ratio is found to be $130:1$ and $100:1$, respectively.


\section{Modeling}
\label{model}

In \cite{ryde&richter} we presented the detection of the $12\,\micron$ \mgi\
emission from Procyon and briefly described our modeling of the emission.
Here, we describe the modeling in more detail.

\begin{figure}
\centering
\includegraphics[width=8cm]{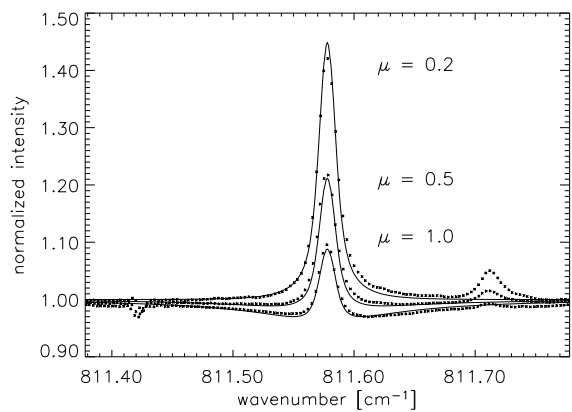}
\caption{The dotted lines show the measured, solar intensity-profiles of the \mgi\ line at $12.32\,\micron$
observed by \cite{brault83}. Three positions on the solar disk are shown, with $\mu = \cos\Theta$ giving
the position on the disk.
The solid lines show our modeled emission lines. The small emission line at $811.71\,$\invcm\  is due to silicon and the
absorption at $811.42\,$\invcm\ is a telluric absorption line.}
   \label{solar_811}
\end{figure}

\begin{figure}
\centering
\includegraphics[width=8cm,height=6.5cm]{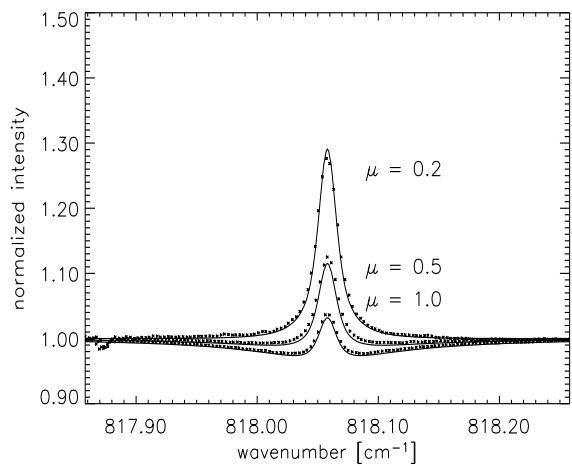}
\caption{The dotted lines depict the \cite{brault83} observations of the $12.22\,\micron$ \mgi\ line
observed from the Sun.   As in Figure~\ref{solar_811}, the model
is shown by  solid lines. The
absorption feature at the very left is a telluric absorption line.}
\label{solar_818}
\end{figure}

We analyze the line formation and the spectral synthesis of the emission
lines with a full non-LTE
calculation using the code MULTI \citep{multi}. We have used
the model atom of \cite{mc_mg}, which was slightly improved by
\cite{bruls:2}, and was kindly provided by Mats Carlsson. To test
our application of the model, we first reproduced the excellent
fits of the intensity profiles measured at different locations on
the solar disk as presented in \cite{mc_mg}. Our model atom is
identical to the one used by \cite{bruls:2}, except that, in order
to fit the solar lines with our modeling, we have increased the
collisional cross-sections for the two $12\,\micron$ lines by
$50\%$, as discussed in Section \ref{increase}. In Figures \ref{solar_811} and
\ref{solar_818} we show the observed intensity profiles of the
solar \mgi\ lines \citep{brault83}, measured at $\mu = \cos \theta
= 0.2$, 0.5, and 1.0 on the solar disk. Our best model is shown by
the solid lines.

The model magnesium atom consists of 66 levels and includes 315 line
transitions and 65 bound-free (bf) transitions from all levels.
The \mgii\ ground state is included.
We solve the equations of statistical equilibrium governing the
level populations for all levels.
A photospheric radiation field is included for calculating
the photo-ionization rates by incorporating the calculated
specific, mean-intensity field for all
depths from a model atmosphere.
This allows full line-blanketing to be considered,
especially in the ultraviolet wavelength
region that affects mainly
photo-ionization from the lowest states.


The physical structure of the atmosphere is modeled with a one-dimensional,
hydrostatic, flux-conserved, non-magnetic,
LTE model-atmosphere, calculated using frequency Opacity Sampling (OS) with the
MARCS code.  MARCS was first developed by \citet{marcs} and has been
successively updated ever since (Gustafsson et al. 2004, in preparation).
Data on absorption by atomic species come from the VALD
database \citep{VALD} and Kurucz (1995, private communication).
Absorption by molecules is included but not very important for Procyon.
The model atmosphere has a plane-parallel geometry and
is horizontally homogeneous,
with 55 depth points
extending out to an optical depth, calculated in the continuum
at $500\,$nm, of $\log\tau_{500} \sim -5$.
No chromosphere is
included.  Our calculation provides the self-consistent, specific,
mean-intensity field for all depths mentioned above.


The fundamental stellar parameters of Procyon, used in the calculation of the model atmosphere,
are based on the discussion in \citet{carlos}: $T_\mathrm{eff} = (6512\pm 50)$~K, $\log g = 3.96\pm0.02$
(cgs units), $M=(1.42\pm0.06)\,M_\odot$, $R=(2.07\pm0.02)\,R_\odot$,
and a metallicity (iron abundance) slightly lower than solar.
\citet{fuhrmann} measures a slightly super-solar Mg
abundance in LTE, whereas \citet{korn_phd} measures a slightly
sub-solar magnesium abundance for a non-LTE analysis. In the
following discussion we will assume a solar abundance mixture as
given by \citet{solar}.

The $12\,\micron$ emission-line parameters are given in Table~\ref{data}.
To fit the width of the lines in Procyon, we need to convolve our
synthetic lines with a Gaussian distribution having a FWHM of $9.5\,$km\,s$^{-1}$.
This includes the atmospheric macroturbulence and projected rotation of the star,
and the contribution of the instrumental profile,
and is somewhat larger than the value \cite{gray:81}
found for Procyon from optical spectra.




\section{Analysis}

\begin{figure}
\centering
\includegraphics[width=8cm,height=6.5cm]{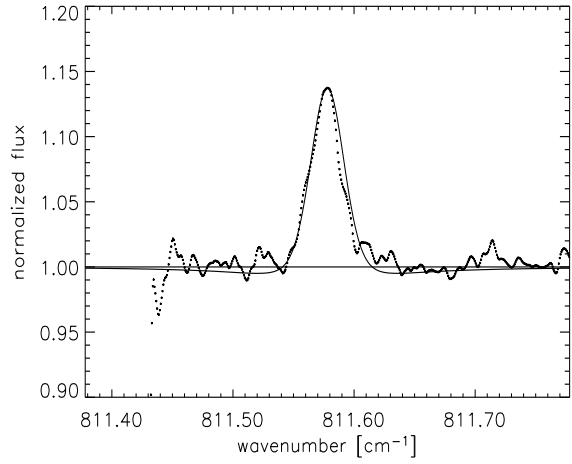}
\caption{The dotted line shows the \mgi\ emission-line at $12.32\,\mic$
observed from Procyon.
The solid line shows the model emission line. A horizontal line showing the normalization level is
added.}
   \label{em_811}
\end{figure}

\begin{figure}
\centering
\includegraphics[width=8cm,height=6.5cm]{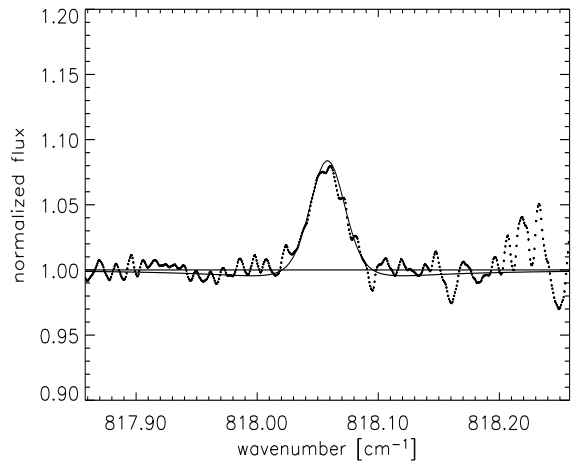}
\caption{The dotted line depicts the $12.22\,\mic$ \mgi\ line observed
from Procyon.
As in Fig~\ref{em_811}, the model is shown by a solid line and
the normalization level is shown by a horizontal line.}
\label{em_818}
\end{figure}

The results of our model for the \mgi\ lines,
$3s7i^{1,3}\mathrm{I}^e \rightarrow 3s6h^{1,3}\mathrm{H}^o$ (811.578\,\invcm)
and $3s7h^{1,3}\mathrm{H}^o \rightarrow 3s6g^{1,3}\mathrm{G}^e$
(818.058 \invcm), in Procyon are shown in
Figures \ref{em_811} and \ref{em_818}, respectively.
The model fits agree well with the observations.
It is interesting to note that the \mgi\ emission in Procyon is of the same
magnitude as the one from the Sun.  As explained in \cite{ryde&richter},
although the higher temperatures in the atmosphere of Procyon
ionize a factor of ten more magnesium atoms,
neutral magnesium is more highly excited.
This compensates for the lower number of atoms and gives rise
to a similar number of effective emitters/absorbers.

\begin{figure}
\centering
\includegraphics[width=8cm,height=6.5cm]{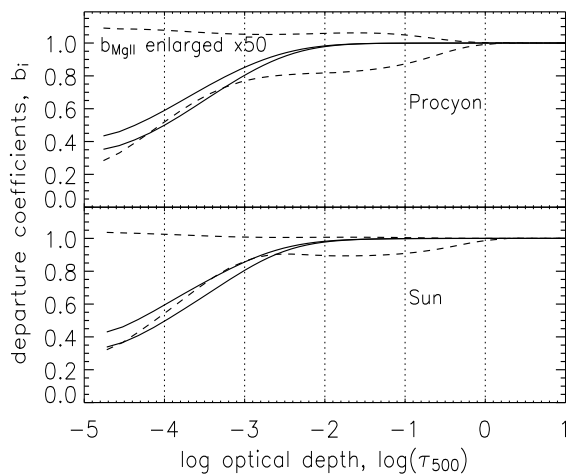}
\caption{Departure coefficients as a function of optical depth
at 500 nm ($\log \tau_{500}$) for the levels involved in the
811.6 and 818.1 \invcm\ lines are shown with full lines. The upper
panel shows our calculations for Procyon and the lower panel shows
the values for the Sun.
The upper and lower levels of the two lines behave in the same way, but
the lower levels depart more from LTE.
The upper dashed line shows
the departure coefficient of the \mgii\ ground state, showing
a larger magnitude for the Sun.
The lower dashed line shows the ground state of \mgi.}
\label{dep_coeff}
\end{figure}

In Figure \ref{dep_coeff}, we have plotted the departures from
Boltzmann level-populations yielded by our non-LTE calculations,
both for Procyon and for the Sun, as a function of optical depth.
We show only the levels involved in the formation of the \mgi\
emission lines and the ground states of \mgi\ and \mgii.  The
departure coefficients are
very similar for both stars, indicating that the same formation
process is at play. Also, they are quite similar for both the
upper and lower states; the departure coefficient of the upper
level is of the order of 10\% larger than that of the lower state
outside of $\log \tau_{500}\sim -2.5$. Thus, the lower levels
depart more from LTE. Whereas for the Sun the population of \mgii\
is slightly larger than that of LTE, for Procyon the \mgii\
population is very close to LTE. In Figure~\ref{dep_coeff}, the
departure coefficient of \mgii\ in Procyon is enlarged.


The small difference between the level departures generating the lines
has a large impact on the emergent line strengths and is the direct cause
of the observed emission.
As also explained in detail in \cite{mc_mg}, the main cause of the
emission is the divergence of the
departure coefficients of the two levels, which
starts to be pronounced at a depth of
$\log \tau_{500}\sim-2$ and increases outwards through the atmosphere.
Although small, the divergence causes the line source function to grow
rapidly outwards in the atmosphere, a fact that increases
with wavelength (due to the increasing influence of
stimulated emission\footnote{In passing, it can be noted that the emission is, however, not
caused by truly inverted populations as is the case in a laser.}). The rapid growth of the source function outwards results in the emission lines.
To a lesser degree, the line opacity is also
affected by the departures.


The departure
coefficients of the upper and lower states, respectively,
of the
two \mgi\ emission lines
follow each other exactly
since they have the same \emph{n} quantum numbers. The reason for this is
our inclusion of (following Carlsson et al. 1992b\nocite{mc_mg}),
`quasi-elastic \emph{l}-changing'\footnote
{\emph{l} being the azimuthal quantum number}
collisions with neutral particles, which
keep all close-lying Rydberg states with common principal quantum numbers, \emph{n}, in
relative thermalization.

The average height of formation of the continuum lies at
approximately $\log \tau_{500}=-1.6$, where H$^-_\mathrm{ff}$
dominates as the continuous-opacity source. The average height of
formation of the line center, calculated in non-LTE, lies at $\log
\tau_{500}\sim-2.7$, but the lines have contribution functions
extending over approximately $-4 < \log \tau_{500} < -1$. The
monochromatic source function at the wavelengths of the line
center decouples from the thermalized Planck-function around $\log
\tau_{500}\sim-1.5$. The line profile probes the minimum of the
source function which causes the troughs of the wings of the observed
line profiles. These are especially prominent in the intensity
profiles in the Sun, see Figures \ref{solar_811} and
\ref{solar_818}.
The troughs are also present in our
modeled line-profiles of the flux from Procyon, but they are too shallow to be
verified in the observed profiles (see Figures \ref{em_811} and \ref{em_818}).


The observed emission lines are formed as a result
of a non-LTE flow cycle (`photon suction', see Carlsson et al.).
The process is driven by lines originating from levels of high-excitation
energy ($\chi_\mathrm{exc} \sim 7$ eV) that become
optically thin in the photosphere.
This leads to over-populated levels at $\chi_\mathrm{exc} \sim 6$ eV, that
are subsequently
photoionized directly or through a state at $\sim 3$ eV.
Note that in the case of the Sun, more than 93\% of the magnesium is
singly ionized, in Procyon more than 99\%.
Through recombinations, this \mgii\ reservoir
eventually refills the levels at $\chi_\mathrm{exc} \sim 7$ eV
through a collisionally coupled ladder of Rydberg states of \mgi.
The emission of the $12~\,\micron$
lines originates from levels next above the states of
the optically thin, $\chi_\mathrm{exc} \sim 7$ eV lines.
It should be noted that there is a strong collisional coupling between
Rydberg states with $\Delta n = 1$, which ensures that the replenishment
proceeds stepwise down through the Rydberg states \citep{rutten:154}.


\citet{avrett:154} find that the solar, $12\,\micron$ emission
is sensitive to collision and photo-ionization rates, but quite
insensitive to the atmospheric parameters.
The atmospheric parameters we use for Procyon
are quite well determined (see Section \ref{model}).
In the following sections, we will investigate our modeling by discussing
the ionizing radiation field, the assumption of homogeneity,
the atomic model, and the collisional data.

\subsection{The ionizing UV field}

\begin{figure*}
\centering
\includegraphics{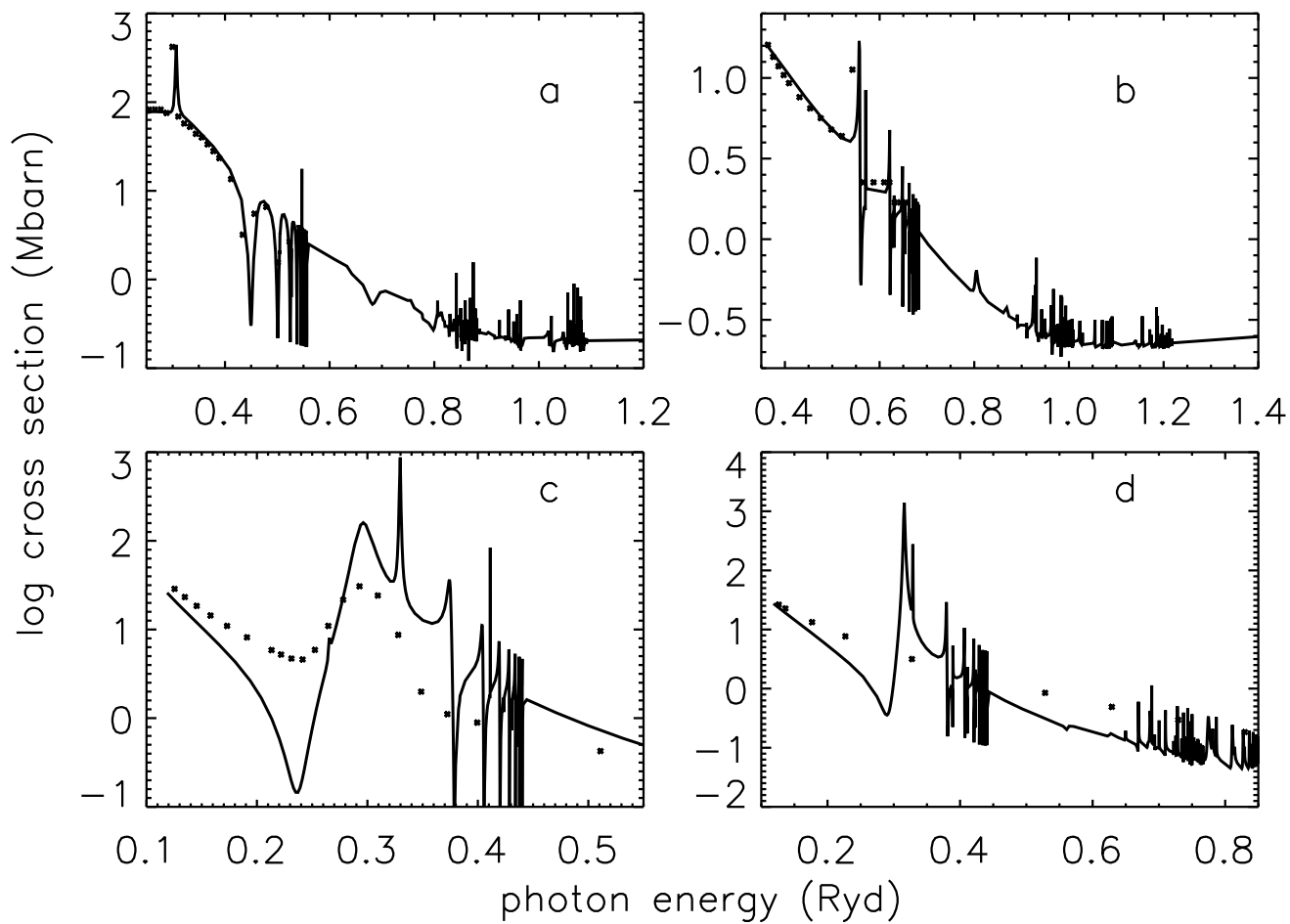}
\caption{Photo-ionization cross-sections from the Carlsson et al. (1992b)
atom are shown by dots
and those from the Opacity Project database \citep{butler} are shown by full lines. The four panels show the four most
influential transitions, in decreasing importance: \textbf{a} shows the ionization from
$3s3p^{1}\mathrm{P}^o$, \textbf{b} shows  $3s3p^{3}\mathrm{P}^o \rightarrow$ \mgii
, \textbf{c} shows
$3s3d^{3}\mathrm{D}^e \rightarrow$ \mgii, and \textbf{d} shows  $3s4p^{3}\mathrm{P}^o \rightarrow$ \mgii .
}
\label{photo_ion}
\end{figure*}

The treatment of the ionizing UV field is of great importance since,
in general, the main non-LTE effect in magnesium is caused by the
photo-ionization, and it affects the strengths
of the \mgi\ lines at $12\,\micron$ \citep{bruls:2}.
The UV field influences the ionization stage of magnesium, and, more
importantly, the photo-ionization from  low-lying levels
closes the non-LTE flow and populates the Rydberg levels.
The ionization state is determined mainly by the photo-ionization
and radiative-recombination equilibria of the three lowest levels:
$3s3s ^1$S$^e$ (with an ionization edge at $1621$\,\AA ),
$3s3p ^3$P$^o$ ($2513$\,\AA ), and $3s3p ^1$P$^o$ ($3756$\,\AA ).
The radiative cross-sections
to and from these levels are well-determined \citep[see, for example,][ or panels a and b
in Figure \ref{photo_ion}]{mc_mg}.

To investigate the sensitivity of the $12\,\micron$ emission lines
to the UV field,
we simulate a change in the UV field by artificially manipulating
the photo-ionization cross-section from these three lowest states.
This can be justified since the photo-ionization rate per unit volume
is an integral of the product between the cross section and the mean intensity (i.e. `UV field' for the
three lowest cross sections) and is  given by
\begin{equation}
\mathrm{Rate} = 4\pi \int_{\nu_o}^\infty \frac{\sigma(\nu) \times J_\nu}{h\nu}\,\,\,\rm d\nu,
\end{equation}
with $\sigma(\nu)$ being the monochromatic cross-section per
particle, $J_\nu$ the mean intensity, and $\nu_0$ the threshold
frequency. For the ground state $3s3s ^1$S$^e$, with an ionization
edge at $1621$\,\AA , a change by as much as a factor of ten does not affect the
emergent intensity at $12\,\micron$ at all. Therefore, the
calculated flux and the modeled photo-ionization cross-section at
that ionization edge was not important for the ionization balance.
This behavior is likely to be due to the fact that the stellar flux at this edge being so
low that it does not affect the ionization balance.
Changing the cross-sections
for the other two low-lying levels by a factor of two changes the
strengths of the emission lines by by less than 10\% and 25\%, respectively.


One way of testing the  ionizing flux, is to compare the modeled,
emergent UV flux with satellite measurements.
This will not ensure that
the calculated flux is correct at all depths, and especially at depths where
the ionization takes place, but it will give an indication of how
appropriate the modeled fluxes are.
The most important fluxes are
the ones at the ionization edges, partly since the bf cross-section
decreases rapidly away
from the edge, and partly due to the fact that the ionizing, mean-intensity field decreases rapidly in the Wien
limit of the star's spectral energy distribution; the measured flux at 1621 \AA\ is, for example, a
factor of 50 times lower than the flux at 2513 \AA.
Thus, we are most interested in testing the fluxes around $2513$\,\AA\  and $3756$\,\AA .

\begin{figure*}
\centering
\includegraphics[bb=50 360 600 700,clip]{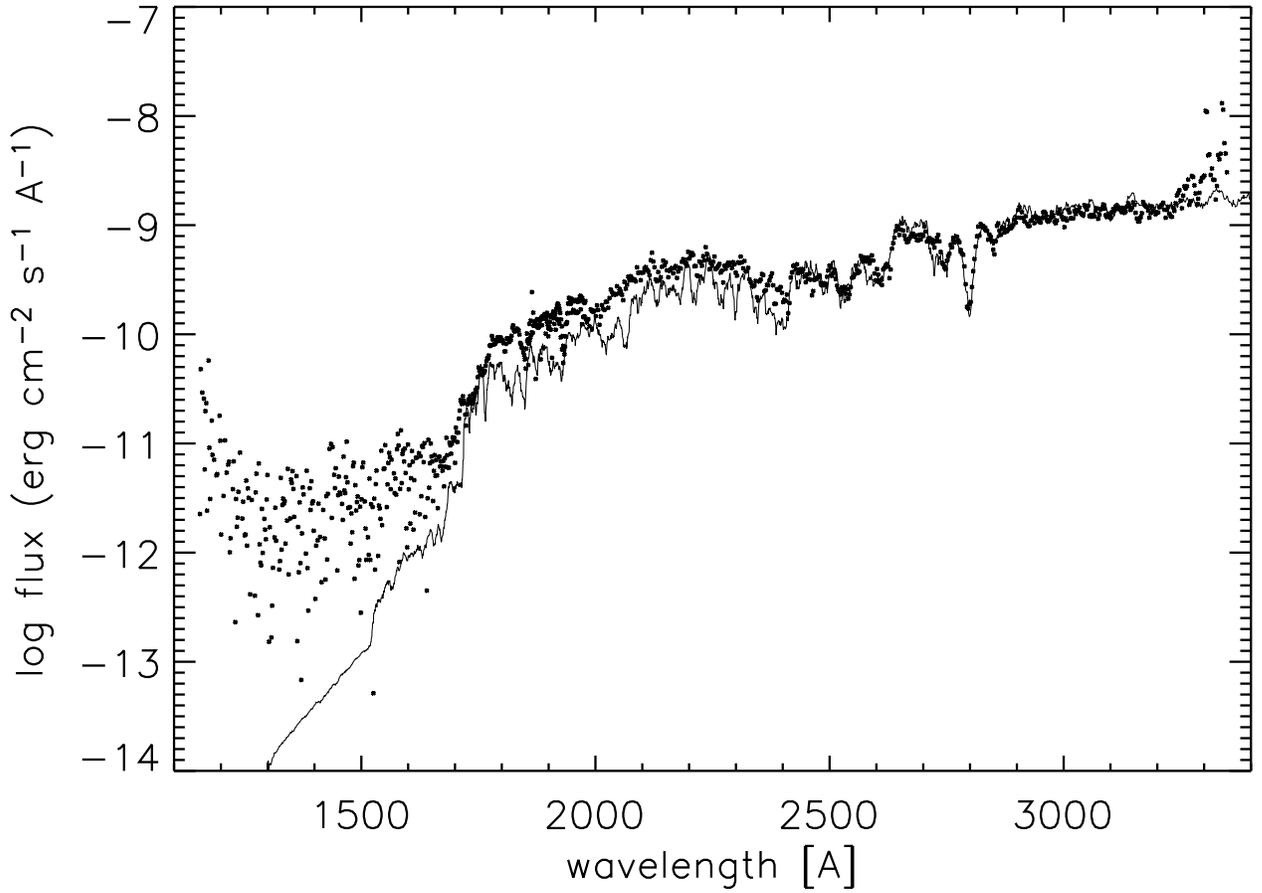}
\caption{Dotted lines show the UV-flux measured by the IUE.
The full lines show the flux we model with the MARCS model-atmosphere code taking into account
a distance to Procyon of 3.5 pc (calculated from a Hipparcos-parallax of 286 mas) and the assumed radius of
$2.1 \,R_\odot$.
}
   \label{iue_model}
\end{figure*}

The International UV Explorer (IUE) satellite\footnote{Based on INES data from the IUE satellite.}
recorded UV spectra for a large number of stars including Procyon.
The SWP spectrometer covered the $1150-1980\,$\AA\ range and
the LWR spectrometer covered the $1850-3350\,$\AA\ region. Correcting for the
distance to Procyon and assuming the above-mentioned radius of the star, it is possible
to compare our modeled spectrum in the UV
with the low-resolution IUE spectrum of Procyon. This is shown in Figure \ref{iue_model}, where the
fluxes are shown on a logarithmic scale.
The modeled flux is the low-resolution, OS-spectrum generated directly from
the MARCS code. The IUE data files,
retrieved from the Archive Data Service (INES), are SWP43428 and LWR09108.
The oval aperture ($10\arcsec\times20\arcsec$)
of the
telescope on-board the satellite ensures
that all flux from the star is recorded. We find that the emergent
flux ($F_\lambda$) from the model
is in fair agreement with the observations. At longer wavelengths, including the interesting
edge at $2513$\,\AA , the agreement is very good with differences within 25\%, see
Figure \ref{iue_model}.
As the flux increases at longer wavelengths, the agreement can be expected to
be even better (e.g. for
the ionization edge at $3756$\,\AA).
For the wavelength range $1800-2400$ \AA, the modeled fluxes
are within a factor of two, and for shorter wavelengths there are
larger discrepancies.

We therefore conclude that our modeled fluxes can be used with
some confidence. The difference between the observed IUE flux and
the modeled flux, of importance for the edges at $2513$\,\AA\  and
close to $3756$\,\AA, is not large enough to affect the modeled
$12\,\mic$ fluxes appreciably.





\subsection{Atmospheric homogeneity}


The granulation pattern in Procyon is shallower than for the Sun.
According to the simulations of \cite{carlos},
the layers in Procyon showing maximum temperature contrast caused
by convection lie where the optical continuum is formed,
$0<\log \tau_{500}<0.5$ (`naked granulation').
For the Sun the corresponding layers lie
beneath the continuum-forming layers ($\log \tau_{500}\sim 1$).
The solar emission lines are not expected to be affected
much by asymmetries caused by granules \citep{rutten:154}.
The continuum is formed so much further out in the solar photosphere that
the line formation is not affected greatly by the
inhomogeneities. Therefore, a plane-parallel, homogeneous model photosphere
is able to reproduce the solar emission lines well.
The \mgi\ emission lines in Procyon have contribution functions
extending over depths of
approximately $-4 < \log \tau_{500} < -1$, but could be affected.

A treatment of the convection in the atmosphere in a more realistic
way would have been preferred but requires atmospheric
models including
3D, hydrodynamic modeling of the convection \citep{asplund_3D,carlos}.
These sorts of models are, as yet, not readily available.
An assessment of  the uncertainties introduced by using traditional models,
which describe convection by
the mixing-length approximation, is given by \cite{carlos}.
They studied Procyon and scrutinized the differences in using
traditional and hydrodynamic models. For example, the difference in
the abundance of iron is found to be less than 0.05 dex.

\subsection{The radiative part of the atomic model}

The atomic model data consist of energy levels, statistical weights,
oscillator strengths ($f_\mathrm{osc}$), broadening parameters (for natural,
van der Waal, and Stark broadening), and photo-ionization rates.
Furthermore, collisional data for line transitions and collisional
ionization are included.
The data used in our calculation are described in \cite{mc_mg}.
We find that the sensitivity of the model emission lines
to the atomic data in the model atom
are of the same kind and order as was found in \cite{mc_mg}, i.e.
a factor of two change in any of the radiative  rates
results in a change of the emission by less than $10\%$, except for the
transition itself. The collisional rates are less influential.


After our model was set up by \cite{mc_mg}, newer
data have appeared in the literature, especially concerning bound-bound
and bound-free radiative transitions, i.e.
$f_\mathrm{osc}$-values, photo-ionization cross-sections, and broadening parameters.
Therefore, we have compared our data for the most important
transitions affecting the \mgi\ emission lines
with the data available at present
from the Opacity Project \citep[OP, ][]{butler}.

The OP data include only levels and transitions for \mgi\ states
with quantum numbers $n < 9$ and $l < 5$.
\cite{zhao} and \cite{przybilla} write that the transition
probabilities from the OP data are accurate to within $\pm 10\%$.
We find that our $gf_\mathrm{osc}$-values are in agreement with the OP data to
within $10\%$ and for most cases to within $5\%$.
For the $12\,\micron$ transitions, our $gf_\mathrm{osc}$-values
differ from those used by \cite{zhao} by approximately $50-60\%$.
While our $gf_\mathrm{osc}$-values come from \citet{mc_mg} and are calculated based on
\citet{green}, the $gf_\mathrm{osc}$-values used
by \citet{zhao} are calculated based on \citet{bates}.
Note that the $f_\mathrm{osc}$-values enter into the calculations of the
collisional rates, thus affecting them by the same amount.


In the \citet{mc_mg} atom, the photo-ionization cross-sections are
given in a smoothed fashion with some of the most important fine
structure, such as resonances, taken into account.
In Figure \ref{photo_ion} we show the most recent OP data for the
photo-ionization cross-sections for the most important transitions
affecting the $12\,\micron$ emission lines.
\cite{przybilla} assume the uncertainty in the OP photo-ionization
data to be of the order of $\pm10\%$.
The two most
influential cross sections are in good agreement with
the OP data. The other two differ more, especially at higher photon energies,
i.e. the further away one gets from the ionization edge.
As these ionization channels close the non-LTE flow, they could be of
importance for the strengths of the emission lines.
Therefore, we repeated our non-LTE calculation using the OP data instead of
the smoothed cross-sections used in the \cite{mc_mg}
atom for these transitions.
We find this hardly affects the emission at all.


New data on Stark broadening due to impacts of charged
particles on \mgi\ have been
published by \cite{dimi}.
Rydberg atoms are physically large and quadratic
Stark-broadening of Rydberg transitions can be
important \cite[see, for example, ][]{mc_mg}.
\cite{dimi} show that the \citet{mc_mg} choice of Stark
broadening is a factor of 2.5 too low compared to new
calculations.
Recalculating our model of the \mgi\ lines with the appropriate
Stark broadening factor,
we find little change.


To summarize, we are confident that the radiative parameters
of the \mgi\ atom presented by \citet{mc_mg}, and which we use, is
up-to-date and performs well for the
\mgi\ emission lines.

\subsection{Collisional data}\label{increase}

In the model atom provided by \citet{mc_mg}, transitions caused by
collisions with electrons are incorporated mostly using the
impact-parameter approximation of \citet{seaton:62}. This provides
a  consistent set of collisional rates accurate to within a factor
of two \citep{mc_mg}.  Rates from the ground state to a few of the
lowest states are taken from \cite{mauas}. As discussed by
\citet{mc_mg} and \citet{bruls:2}, the most important aspect of
the collisional data is its homogeneity throughout the atom.
The internal consistency is significant for the set-up of the non-LTE
populations throughout the atom and, therefore, for the formation
of the $12 \,\micron$ lines.


From our non-LTE calculation we see that the collisional rates are
larger than the radiative rates for the levels involved in the
generation of the emission lines by a factor of 3 at line center and
a factor of 50 at the extremes of the line wings.
The fact that collisions dominate the Rydberg transitions was also
found for the solar case by \citet{MC_csss}.
The collisions play an important role in setting up the
Rydberg ladder through which the recombined atoms de-excite.
In order to reproduce the fits of the solar lines,
we have increased the collisional cross-sections for the
two $12\,\micron$ lines by $50\%$ compared to the
ones given by the impact approximation, which is well within the uncertainties.
Finally, we also find that changing the collisional \emph{ionization}
by as much as a factor of ten did not result in any noticeable difference.


The \cite{zhao} non-LTE model includes
inelastic, allowed neutral-hydrogen collisions, which should be
increasingly important for cool, metal-poor stars.
The strength and energy dependence of these collisions were
empirically determined.
The authors show that including the collisions influences
the line-cores of optical and
near-infrared \mgi\ lines by
reducing the non-LTE effects by approximately
50\% and results in better fits to the solar \mgi\ lines they considered.
The line cores are formed
at shallow depths in the atmosphere where electron collisions are less
important and where hydrogen collisions can be expected to be important.
\cite{zhao} incorporate the collision rates with the \cite{drawin} formula.
These rates are, however, subsequently scaled with a factor that decreases
exponentially with the excitation energy of the upper level involved in
the collision, thus not affecting the $12\,\micron$ \mgi\ emission lines.
We chose not to include these extra collisions, since their
global effect on
the departure coefficients of the \mgi\ atom is assumed to be small.
The hydrogen collisions can, however, have an effect on the magnesium abundance
inferred from optical lines which usually enters the analysis of the
Mg I IR emission lines as an additional unknown. However, as mentioned above, in our analysis of
Procyon we have assumed solar composition.

\subsection{Further observations}

\begin{figure}
\centering
\includegraphics[width=8cm,height=6.5cm]{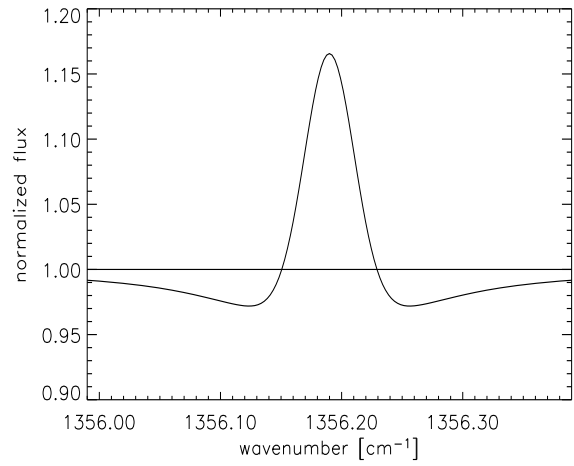}
\caption{Prediction of the strong \mgi\ emission-line at
$1356.2$~\invcm . It is plotted on the same relative scale as in
Figures \ref{em_811} and \ref{em_818}, making a direct comparison
possible. The troughs in the flux spectrum of this line are
pronounced and should be detectable at a signal-to-noise ratio
similar to our present observations of the other two \mgi\ lines.
}
   \label{1356}
\end{figure}

%
With our model we find that a stronger line is
at $1356.19$\,\invcm, generated by the $6h^{1,3}\mathrm{H}^o
\rightarrow 5g^{1,3}\mathrm{G}^e$ transition of neutral magnesium.
The model prediction for Procyon is shown in Figure \ref{1356}.
The troughs are more
pronounced than those of the 811.6 and 818.1 \invcm\ lines, and
should be detectable. They are of interest since they map the
minimum of the source function as a function of depth in the
atmosphere.
The troughs are very broad, of the order of $\sim$75~km\,s$^{-1}$
and will require care when
normalizing the spectra to preserve the troughs during data reduction.
Ground-based observations of this line require very low column density of
telluric water vapor,
but future observations from the Stratospheric Observatory
for Infrared Astronomy (SOFIA) will be fairly routine.
Because molecular transitions of H$_2$O, SiO, and CO are common throughout
this region, we expect the $1356.19$\,\invcm\ line to be best for stars
hotter than spectral type M.

\section{Discussion and conclusions}

The splitting of the solar $12\,\micron$ lines in magnetically active
surface regions shows normal Zeeman triplet splitting with
one $\pi$ and a pair of $\sigma$ components, although all
the components actually have tightly grouped substructure.
The $\pi$ and $\sigma$ components are separated by a width given
by an effective Land\'e factor of approximately 1,
which was experimentally found
by \cite{lemoine}.
The average separation between the two $\sigma$ components is given by
\begin{equation}
\label{ekvation}
     \Delta\lambda_B = 2\times 4.67\times10^{-9}g\lambda^2_0 B \,\,\,(\mu\mathrm m ),
\end{equation}
where $g$ is the Land\'e factor, $\lambda_0$ is the
unperturbed wavelength of the line in $\micron$, and $B$ the magnetic field
strength in Gauss (G).
The strength of the solar magnetic fields puts
the Zeeman effect in the Paschen-Back regime for these lines \citep{bruls:2}.

%

There are advantages to Zeeman magnetic-field studies using infrared lines
compared to optical ones.
In addition to less line-blending due to the lower line density \citep[see, for example, ][]{ryde_munchen},
infrared spectral lines have larger Zeeman sensitivity.
Optical lines most often do not show large enough Zeeman separation in order to be disentangled from other
broadening mechanisms of the line.
The ratio of the separation of the Zeeman splitting ($\Delta\lambda_\mathrm{B}\sim g\lambda^2B$) and the non-magnetic
Doppler width
($\Delta\lambda_\mathrm{D}\sim\lambda$), is given by
$\Delta\lambda_\mathrm{B}/\Delta\lambda_\mathrm{D}\sim g\lambda B$.
Thus, the ratio grows with wavelength and,
consequently, it is easier to detect Zeeman
split lines in the infrared. With larger
Zeeman separation between the components, the analysis is
simpler.


Although the largest Land\'e factors for optical and near-infrared transitions
are roughly 3 while the
\mgi\ lines at $12\,\micron$ have an effective Land\'e factor of 1,
moving to $12\,\micron$ from the
optical and near-infrared, respectively, increases the sensitivity by
at least a factor of 8 and 3, respectively.
According to \citet{solanki:154}, it should be possible to measure
solar magnetic field strengths as low as $150-200$ Gauss from line
splitting, and fields down to
100 Gauss are  measurable through (very) careful profile fits.
(A large magnetic field would split the line and in the case of
low field-strength it would act as an additional broadening.)
\cite{bruls:2} discuss measurements of the Stokes parameters of
the solar $12\,\micron$ lines affected by solar-magnetic fields,
and indicate that field-strengths of approximately 200 G are
measurable through \emph{Stokes V}-line splitting. They also show
that line-splitting in the \emph{Stokes I} light is reliable only
for strengths greater than 500 G.

It should be noted that, especially for the broadened lines, the
sensitivity for detecting magnetic fields for a disk-averaged
stellar, \mgi\ spectrum will also depend on the filling factor
($f$) of the field over a stellar surface \citep[see for example,
][]{valenti95}. In the simplest case, a stellar magnetic field can
be modeled by a two-component model, with one magnetic component
with a constant field strength, $B$, over a fraction, $f$, of the
surface and one non-magnetic component, the latter covering
($1-f$) of the stellar surface \citep{solanki:invited}.
Furthermore, in a field-free approximation \citep[see, for
example, ][]{bruls:2} one can assume that a magnetic field does
not affect the departure coefficients significantly. The non-LTE
calculations can, therefore, be made in a non-magnetic environment
(Piskunov, private communications).

We have not resolved any magnetic components on Procyon and see only
single peaked features that are well fitted by Gaussian distributions.  Given the
absence of any resolved $\sigma$ components, we can estimate the field
strength and filling factor that would have produced detectable,
separate, $\sigma$ components.
We consider the line at 811.578~\invcm\ because the signal-to-noise
ratio for this line, 130:1, is higher than for the 818.058~\invcm\ line.
Our estimate considers only the magnetic field strength and filling
factor at the photospheric height that dominates the line formation.
Pressure equilibrium and magnetic flux conservation suggest that
magnetic flux tubes will have a larger filling factor  and weaker
field strength higher in the
photosphere where the gas pressure is lower.

Given our observations, we believe we would have detected resolved
$\sigma$ components if the magnetic field strength had been $B>830\,$G
and the filling factor $f>26\%$.  In this simplistic estimate, we
assume the observed line, which has a peak height of 0.135
above the normalized continuum, comes from the non-magnetic region with filling factor
$(1-f)$.  Given the signal-to-noise ratio, a 3$\sigma$ detection would
require a peak value of 0.024 above the normalized continuum.
If the efficiency for producing the two $\sigma$ components in the
magnetic region is the same as for the central peak from the non-magnetic
region, we can calculate that the magnetic filling factor must be approximately  26\%.
To determine the maximum field strength allowed by our observations,
we consider a separation between the two $\sigma$ components of
28.5~km\,s$^{-1}$ (3 times the FWHM of the detected line)
to be sufficiently resolved that we would resolve
the separate peaks.  Using Eq. \ref{ekvation} given above, such a  splitting
would require a magnetic field strength of 830~G.
For comparison, it can be noted that similar and also stronger fields
than this have been detected for other stars. For example,
\cite{johnskrull99} measure a surface-averaged mean-field from IR observations of
the T Tauri star BP Tau of $\Sigma Bf=2.6\pm0.3$\,kG,
and \citet{valenti95} find that 9\%\ of the deep photosphere of the active
star $\epsilon$ Eri is filled with a 1.4 kG field.


A further diagnostic value of the $12\,\micron$
lines is their different height of formation compared with lines
at, for example, $1.5\,\micron$.
The $1.5\,\micron$ lines are commonly used in solar, magnetic-field
observations and sample the opacity minimum.
For the Sun, the lateral expansion of magnetic flux tubes with height
dilutes the field strength \citep{deming}. Thus fields measured at
$12\,\micron$, sampling higher regions, may be lower than those
measured by lines lying at $1.5\,\micron$.
Because of the different depth dependence, comparisons of Zeeman
splitting from lines at different frequencies can start to trace the
vertical structure of magnetic flux tubes \citep{bruls:2}.

While the $12\,\micron$ lines provide high sensitivity to magnetic
fields, observing these lines is still difficult. As $8-10$\,m
class telescopes become equipped with high-resolution,
mid-infrared spectrographs, many more stars will prove suitable
targets. We are confident that high-resolution observations of the
$12\,\micron$ lines will, certainly, be an important tool for
measuring \emph{stellar} magnetic fields in the future. We have
shown that we understand the formation of the $12\,\micron$ lines
not only for the Sun but also for Procyon and that the non-LTE
calculation, based on the one for the Sun, is still up-to-date.

\acknowledgments

 We should like to thank Dr. Mats Carlsson for providing the \mgi\ model-atom used in
this study. We gratefully acknowledge our fruitful discussions with
Drs. Kjell Eriksson and Bengt Gustafsson, and thank Karin Ryde for linguistic aid.
We are grateful for the help of Drs. John Lacy and Thomas Greathouse
of the TEXES team as well as the {\sc irtf} staff. This work was
supported in part by the Swedish Research Council, the Swedish Foundation for International Cooperation
in Research and Higher Education, and Stiftelsen Blanceflor
Boncompagni-Ludovisi, n\'ee Bildt. The construction of the TEXES spectrograph was
supported by grants from the {\sc nsf} and observing with TEXES was supported by the Texas Advanced Research Program.
A. J. Korn was supported by the Leopoldina Foundation/Germany under grant
BMBF-LPD 9901/8-87.


\clearpage

  \begin{table*}
\footnotesize{
      \caption[]{Line data of the $12\,\micron$ lines.}
         \label{data}
     $$
         \begin{array}{cccccccc}
            \hline
            \noalign{\smallskip}
            \rm{wavelength}     &  \rm{wavenumber} & \rm{transition} & E_\mathrm{up} & E_\mathrm{low} & {\rm log}\, gf & \mathrm{Lande\,\,g\,\,factor}\\
            \lambda        & \sigma & & & & \rm Green\,\,et\,\,al.\,\, (1959) & \rm Lemoine\,\, et\,\, al.\,\, (1988)\\
            \rm [\AA ]  & \rm [cm^{-1}] & & \rm [eV] & \rm [eV] & \rm (cgs)  \\
            \noalign{\smallskip}
            \hline
            \noalign{\smallskip}
            12.3217     & 811.578 & 3s7i^{1,3}\mathrm{I}^e - 3s6h^{1,3}\mathrm{H}^o & 7.3693 & 7.2686  & 1.95 & 0.99\pm0.01 \\ 
            12.2241     & 818.058 & 3s7h^{1,3}\mathrm{H}^o - 3s6g^{1,3}\mathrm{G}^e & 7.3691 & 7.2676  & 1.73 & 1.00\pm0.01 \\
           \noalign{\smallskip}
            \hline
         \end{array}
     $$}
   \end{table*}

\clearpage

\end{document}